# Quantifying Uncertainty in Infectious Disease Mechanistic Models

Lucy D'Agostino McGowan, Kyra H. Grantz, and Eleanor Murray

Author affiliations: Department of Mathematics and Statistics, Wake Forest University, Winston-Salem, NC, United States (Lucy D'Agostino McGowan), Department of Epidemiology, Johns Hopkins Bloomberg School of Public Health, Baltimore, MD, United States (Kyra H. Grantz), Department of Epidemiology, Boston University, Boston MA, United States (Eleanor Murray).

Correspondence to Dr. Lucy D'Agostino McGowan, Department of Mathematics and Statistics, Wake Forest University, 127 Manchester Hall Box 7388, Winston-Salem, NC 27109, (e-mail: mcgowald@wfu.edu)

# Abstract

This primer describes the statistical uncertainty in mechanistic models and provides R code to quantify it. We begin with an overview of mechanistic models for infectious disease, and then describe the sources of statistical uncertainty in the context of a case study on SARS-CoV-2. We describe the statistical uncertainty as belonging to three categories: data uncertainty, stochastic uncertainty, and structural uncertainty. We demonstrate how to account for each of these via statistical uncertainty measures and sensitivity analyses broadly, as well as in a specific case study on estimating the basic reproductive number, $R_0$, for SARS-CoV-2.

keywords: mechanistic models, statistics, uncertainty, SARS-CoV-2, sensitivity analyses, Monte Carlo simulation, infectious disease modeling

# BACKGROUND

Mechanistic models often refer to models based on a specific *structure*. While these are occasionally portrayed in opposition to *statistical* models, statistical uncertainty is an important aspect of interpreting mechanistic model results. With the current emphasis on using mechanistic models for the COVID-19 pandemic response, having a keen understanding of the uncertainties that underpin these models is particularly important, both for those creating models and those consuming results. Unfortunately, understanding the sources of uncertainty in infectious disease models has been shown to be difficult, even for experts.(1)

This short primer defines three main types of uncertainty that accompany mechanistic models: data uncertainty, stochastic uncertainty, and structural uncertainty. We begin with an overview of infectious disease mechanistic models, and then describe sources of uncertainty in the context of a case study on estimating the expected number of new infections from a single infectious individual, $R_0$, for SARS-CoV-2. We provide R code (R Core Team, 2020) to calculate each component we estimate. We note that mechanistic models are often used for one or both of two purposes: to draw inference about transmission dynamics, including estimating key parameters, and to predict disease transmission and outbreak trajectories. This primer focuses on estimating $R_0$ as an example, because it provides a straightforward demonstration of the uncertainty one may want to quantify, but we also highlight considerations for other use cases. The goal of this primer is to provide the reader with the tools to:

(1) Understand sources of uncertainty in mechanistic models
(2) When consuming literature where mechanistic models are used, be able to identify which sources of uncertainty were accounted for in the intervals presented, and which were not.
(3) Calculate uncertainty parameters using the R code provided

This primer is intended to illustrate through a clear example how one may interrogate an estimate obtained from an infectious disease model to further understand the potential sources of uncertainty and how sensitive the estimate may be to them.

## Mechanistic Modeling

The modeling of disease outbreak trajectories can take many forms, but mathematical mechanistic models are most common.(2) These compartmental models simulate movement of populations or individuals through a series of mutually-exclusive and exhaustive health states over time. They require specification of at least three types of information: states to be modeled, cycle length, and parameters that govern the transition from some states to others. A *state* describes a distinct health status. States must be mutually exclusive and exhaustive but may be defined as broadly or as narrowly as necessary to answer the research question. A *cycle* is one time step. For fixed-time models, the model runs through a decision tree to determine the distribution of states at the next time point. Typical values for a cycle length are one day, one month, or one year. A *transition* occurs when an individual moves from one health state at the start of a given cycle to another health state at the end of that cycle. *Transition parameters* describe the cycle-length-specific probability of moving (transitioning) from one health state to another health state in a given cycle.

*Infectious disease model definitions.* Key terms used in infectious disease models can be categorized as regarding *states* or *transition parameters*. The most common *states* for directly-transmissible diseases are:

(1) Susceptible (S)
(2) Exposed (E)
(3) Infectious (I)

(4) Recovered (R)

Individuals are assumed to occupy only one state at a time. Individuals in the susceptible state (S) are those that have the potential to contract the disease. Individuals in the exposed state (E) have been exposed to the disease but are not yet infectious. Individuals in the infectious state (I) have been infected and *are* infectious, and those in the recovered state (R) have recovered. Individuals in the recovered state are often assumed to remain there due to immunity, however there are model constructions that allow them to move back into the susceptible state once immunity wanes.

Transition parameters describe how individuals move from one state to the next. These include:

(1) $\beta$: The transmission coefficient. This describes the rate at which individuals move *out* of the susceptible state.

(2) $\sigma$: The rate of becoming infectious. This describes the rate at which individuals move *out* of the exposed state.

(3) $\gamma$: The recovery rate. This describes the rate at which individuals move *out* of the infected state.

These parameters then correspond to the duration of each state:

(1) $1/\sigma$: the duration of the latent period

(2) $1/\gamma$: the duration of the infectious period

Two additional important definitions are:

(1) $\lambda$: The growth rate of the epidemic

(2) $R_0$: The expected number of new infections from a single infectious individual in a completely susceptible population, referred to as the basic reproductive number

The calculation of the growth rate of the epidemic, $\lambda$, and the basic reproductive number, $R_0$, depend on the chosen underlying *structure* of the infectious disease mechanistic model. The following sections describe three possible structures.

*SIR Models.* A common model to describe a disease outbreak is an SIR model(3) where individuals transition between three states: susceptible (S), infected (I), and recovered (R) (Figure 1).

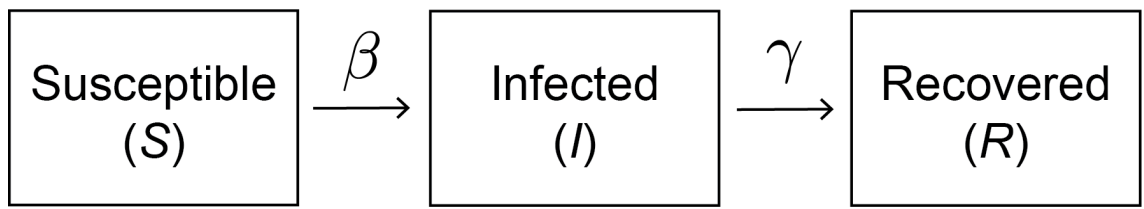

**Figure 1**. SIR model

In this basic formulation, two parameters, $\gamma$ and $\beta$, are specified, the rate at which the infectious recover, and the rate at which potentially infectious contacts are made, respectively. The transmission cycle is then solved via a set of ordinary differential equations (ODEs), differentiated with respect to time (Equation 1).

$$S' = -\frac{\beta IS}{N}$$
$$I' = \frac{\beta IS}{N} - \gamma I \qquad (1)$$
$$R' = \gamma I$$

where ′ indicates the derivative is taken with respect to time, as defined by the cycle length (Box 1), and $N$ is equal to $S + I + R$, the total number in the population at a given time. The duration of the infectious period is defined as $1/\gamma$. This basic model assumes no latent period for the disease and that the infectious period is exponentially distributed. The expected number of new infections from a single infectious individual is referred to as the basic reproductive number, $R_0$, is the ratio between $\beta$ and $\gamma$ (Equation 2).

$$R_0 = \frac{\beta}{\gamma} \qquad (2)$$

This can also be parameterized from the expected exponential growth rate of the epidemic, $\lambda$ (Equation 3).

$$R_0 = 1 + \frac{\lambda}{\gamma} \qquad (3)$$

where $\lambda$ is the growth rate, $\lambda = \beta - \gamma$. (4)

This estimate of $R_0$ can be calculated using the function found in R Code 1.

```
r_sir <- function(lambda, gamma) {
  1 + lambda / gamma
}

r_sir(0.1, 1 / 5)

## [1] 1.5
```

**R Code 1.** Function to estimate $R_0$ under an SIR structure with an example usage with $\lambda = 0.1$ and $1/\gamma = 5$. If we estimated an initial growth rate of the epidemic from reported case data of 0.1 and the infectious period of 5 days, the estimated $R_0$ value under this SIR structure would be 1.5.

*SEIR Models.* The structure of the basic SIR model can be further updated, for example as an SEIR model, where individuals transition between four states: susceptible (S), exposed (E), infected (I), and recovered (R) (Figure 2).

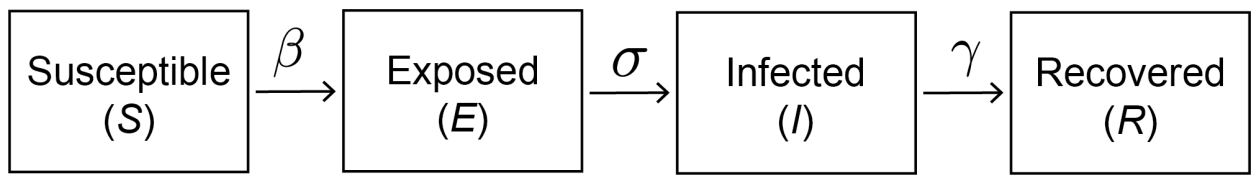

**Figure 2**. SEIR model

The introduction of the additional state, exposed (E), adds a transition parameter $\sigma$ for the rate at which the exposed become infectious. To examine how individuals transition through these states over time, we solve an updated set of ODEs (Equation 4).

$$S' = -\frac{\beta IS}{N}$$
$$E' = \frac{\beta IS}{N} - \sigma E$$
$$I' = \sigma E - \gamma I$$
$$R' = \gamma I$$
(4)

where $N$ is now equal to $S + E + I + R$, the total number in the population at a given time. In this formulation, the duration of the infectious period is still $1/\gamma$. Compared to the SIR model, the duration of the latent period is now non-zero, quantified as $1/\sigma$. Both the latent and infectious periods are assumed to be exponentially distributed. $R_0$ can be expressed the same as an SIR model (Equation 2), or with respect to the expected growth rate, $\lambda$ (Equation 5).

$$R_0 = \left(1 + \frac{\lambda}{\gamma}\right)\left(1 + \frac{\lambda}{\sigma}\right)$$
(5)

where $\lambda$ is now defined as follows in Equation 6.

$$\lambda = \frac{-(\sigma + \gamma) + \sqrt{(\sigma + \gamma)^2 + 4\sigma\beta}}{2}$$
(6)

This estimate of $R_0$ can be calculated using the function found in R Code 2.

```
r_seir <- function(lambda, gamma, sigma) {
  (1 + lambda / gamma) * (1 + lambda / sigma)
}
r_seir(0.1, 1 / 5, 1 / 5)

## [1] 2.25
```

**R Code 2.** Function to estimate $R_0$ under an SEIR structure with an example usage with $\lambda$ = 0.1, $1/\gamma$ = 5, $1/\sigma$ = 5. If we estimated an initial growth rate of 0.1, duration of the infectious period of 5 days, and duration of the latent period of 5 days, we would estimate an $R_0$ value of 2.25 under this SEIR structure.

*SEIR model with relaxed assumptions* One assumption of SEIR models is that the rate of leaving the exposed (E) or infectious (I) class is constant, regardless of how long an individual has been in that class, resulting in the assumption that the latent and infectious periods are exponentially distributed.(5,6) This assumption can be relaxed by introducing additional parameters to estimate the latent and infectious periods using a more flexible gamma distribution. To do this, additional states are added; the exposed (E) state is broken into $m$ substates and the infected (I) state is broken into $n$ substates (Figure 3). The duration of the latent period is estimated using a gamma distribution with shape parameter $m$ and scale parameter $m\sigma$. The duration of the infectious period is estimated using a gamma distribution with shape parameter $n$ and scale parameter $n\gamma$. We refer to these as SE$^m$I$^n$R models to indicate the $m$ exposed states and $n$ infected states.

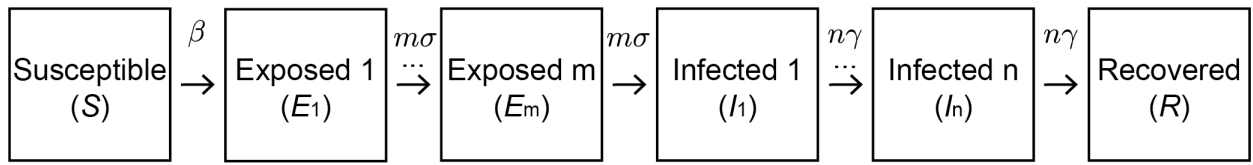

**Figure 3.** SE$^m$I$^n$R model

This results in the ODEs specified in Equation 7.

$$S' = -\frac{\beta \sum_{i=1}^{n} I_i S}{N}$$

$$E'_1 = \frac{\beta \sum_{i=1}^{n} I_i S}{N} - m\sigma E_1$$

$$E'_i = m\sigma E_{i-1} - m\sigma E_i, i = 2,\dots,m$$

$$I'_1 = m\sigma E_m - n\gamma I_1 \tag{7}$$

$$I'_i = n\gamma I_{i-1} - n\gamma I_i, i = 2,\ldots,n$$

$$R' = n\gamma I_n$$

where $N$ is now equal to $S + \sum_{i=1}^{m} E + \sum_{i=1}^{n} I + R$. Note that in this formulation, rather than a single state for E, the exposed and I, infected, there are $m$ exposed states and $n$ infected states. Individuals pass through the exposed states at a rate of $m\sigma$ and through the infected states at a rate of $n\gamma$. The sum of $m$ exponential distributions with a scale parameter of $m\sigma$ results in a gamma distribution with a shape of $m$ and a scale of $m\sigma$. This form reduces to an SEIR model when $m$ and $n$ are both 1, indicating only one state, E, for the exposed and one state, I, for the infected. In the $SE^mI^nR$ formulation, $R_0$ is calculated as specified in Equation 8.(5,7,8)

$$R_0 = \frac{\lambda \left(\frac{\lambda}{\sigma m} + 1\right)^m}{\gamma \left[1 - \left(\frac{\lambda}{\gamma n} + 1\right)^{-n}\right]} \tag{8}$$

where $\lambda$ is the initial growth rate of the epidemic. In this parameterization, the mean infectious period is $1/\gamma$ and the mean latent period is $1/\sigma$.

This estimate of $R_0$ can be calculated using the function found in R Code 3.

```
r_seminr <- function(lambda, gamma, sigma, m, n) {
  (lambda * ((lambda / (sigma * m)) + 1)^m) /
    (gamma * (1 - (lambda / (gamma * n) + 1)^(-n)))
}
r_seminr(0.1, 1 / 5, 1 / 5, 4.5, 3)

## [1] 2.17
```

**R Code 3.** Function to estimate $R_0$ under an $SE^mI^nR$ structure with an example usage with $\lambda$ = 0.1, $1/\gamma$ = 5, $1/\sigma$ = 5, $m = 4.5$, $n = 3$. If we estimated an initial growth rate of 0.1, duration of infectious period from a gamma distribution with a shape parameter of 3, and scale parameter of 0.6 (giving an average infectious period of 5 days), and duration of the latent period from a gamma distribution with a shape parameter of 4.5 and scale parameter of 0.9 (giving an average latent period of 5 days), we would estimate an $R_0$ value of 2.17 under this $SE^mI^nR$ structure.

## Quantifying Uncertainty

Mechanistic models, as used in infectious disease epidemiology, inherently have uncertainty that can be quantified using statistical methods. How this uncertainty is quantified depends on the exact use and desired output of the model. In some cases, the ordinary differential equations specified above can be used to derive closed form solutions for estimating a quantity of interest when other model parameters are fixed; in these situations, uncertainty in the fixed parameters can be incorporated numerically or via simulation. Often, multiple parameters are estimated at once by fitting the model to data on outcomes (e.g., number of observed infections). These methods typically consider a possible distribution of observed outcomes in a likelihood function, though more sophisticated methods exist (9). Simulations are also critical when the goal is forecasting or predicting disease burden or outbreak trajectories, rather than point estimates of single quantities.

There are three commonly conceptualized categories of uncertainty to consider in the generation and interpretation of modeling results, which we discuss in more detail below:

(1) data uncertainty: uncertainty in specified model parameters, estimated externally from data, or in data to which models are fit

(2) stochastic uncertainty: uncertainty derived from the method of simulation; this stochasticity can be of inherent interest (e.g., process stochasticity in transmission events) or an artifact of statistical estimation

(3) structural uncertainty: uncertainty in the optimal model structure, or derived from the use of more than one model structure for a given question

To illustrate these, we discuss their application to a recent report by Sanche et al.(10) The authors used an SE$^m$I$^n$R model, as specified in Figure 3. Under this model structure, they estimate multiple quantities, but we focus on their estimate of $R_0$ of SARS-CoV-2 assuming a serial interval of 7-8 days, demonstrating how this estimate varies as uncertainty is introduced.

# DATA UNCERTAINTY

*Data uncertainty* can generally arise in one of two ways: because a value itself arises from a distribution or from our incomplete knowledge or observation of the true value. For example, most transition parameters, including the latent period and duration of infectiousness, are assumed to naturally follow some distribution in the population; our ability to accurately estimate this distribution, or even the mean or median of a given parameter, may be limited by the number or kinds of observations available.

This uncertainty is the most commonly reported uncertainty in infectious disease models.(6,11,12) In the following section we describe how to quantify the data uncertainty in the context of the Sanche et al. report.

The transition parameters are an integral part of mechanistic models. These are often chosen using available data, and therefore inherently contain a level of uncertainty. This can be handled by assuming parameters arise from a distribution, sampling from the given distribution $M$ times, and calculating a percentile confidence interval (CI) for estimates derived from this sampling procedure.(13) For example, in the simple SIR model (Figure 1), to calculate $R_0$ from its closed form solution derived from the system of ODEs above, we require two parameters: $\lambda$, the exponential growth rate of the epidemic, and $1/\gamma$, the infectious period. A Monte Carlo simulation consisting of $M$ runs, generates $M$ data samples, $\boldsymbol{X} = \{X_1, X_2, \cdots, X_M\}$. Each data sample $X_i$ consists of two estimates, $\lambda$, sampled from its specified distribution $\lambda \sim f$, and $1/\gamma$, sampled from it's specified distribution, $1/\gamma \sim g$. Here, the target quantity of interest, $\phi$, is $R_0$ as specified in Equation 3, estimated as $\hat{\phi}(X)$. The 95% confidence interval is then calculated by ordering the $M$ estimates, $\hat{\phi}(X_1), \cdots, \hat{\phi}(X_M)$, and selecting the 2.5th percentile and 97.5th percentile. We can do this in R using the function specified in R Code 1. Now, instead of assuming that $\lambda$ and $\gamma$ are fixed, as in the R Code 1 example, we will assume they come from a *distribution* of values. There have been several estimates of these values in the literature, for example an analysis from Wuhan, China estimates the growth rate, $\lambda$, to be 0.1 (95% CI: 0.05 - 0.16) (14). For the infectious period, rather than assuming the fixed 5 days, as was done in R Code 1, we could consider a range of values, for example between 4 and 6 days. In the simplest construction, we can assume these come from a *uniform* distribution with a range from (0.05 - 0.16) for $\lambda$ and (4 - 6) for $1/\gamma$. We can simulate this $M = 10^6$ times (R Code 4).

```
set.seed(8) # set a seed so the result is reproducible
M <- 10^6 # set the number of simulations
lambda <- runif(M, min = 0.05, max = 0.16) # simulate initial growth rate
gamma <- 1 / runif(M, min = 4, max = 6) # simulate the infectious period
r0 <- r_sir(lambda = lambda, gamma = gamma) # calculate R0
quantile(r0, c(0.025, 0.5, 0.975)) # calculate the median and 95% CI

##     2.5%     50%   97.5%
##    1.249   1.518   1.859
```

**R Code 4.** Estimating $R_0$ under an SIR model with parameter uncertainty with $\lambda$ from a uniform distribution between 0.05 and 0.16, and $1/\gamma$ from a uniform distribution between 4 and 6.

If we estimated the initial growth rate was uniformly distributed between 0.05 and 0.16, and the duration of the infectious period was uniformly distributed between 4 and 5 days, we would estimate an $R_0$ value of 1.5 (95% CI: 1.2 - 1.9) under this SIR structure. While this example demonstrates how uncertainty can be incorporated in parameter estimation, it is a rather *simplistic* structure. For example, the assumption that the infectious period is uniformly distributed is likely untrue, and this simple structure does not account for an exposed compartment or latent period. Let's turn to a more realistic example, using the Sanche et al. report with an SE$^m$I$^n$R structure.

The recent report by Sanche et al. estimates $R_0$ of SARS-CoV-2 assuming a serial interval of 7-8 days to be 5.8 (95% CI: 4.4 - 7.7).(10) To do so, they use data to estimate the unknown quantities, drawing each from a uniform distribution. The SE$^m$I$^n$R model (Figure 3) used by the authors involves three unknown transition parameters: $\lambda, 1/\sigma$, and $1/\gamma$. They draw each of these parameters from uniform distributions as specified in Table 1.

**Table 1**. *Distribution of transition parameters*[a]

| Transition parameter | Assumed distribution |
|---|---|
| $\lambda$: the growth rate of the epidemic | Uniform distribution between 0.21 and 0.30 per day |
| $1/\sigma$: the mean latent period | Uniform distribution between 2.2 and 6 days |
| $1/\gamma$: the mean infectious period | Uniform distribution between 4 and 14 days |

[a]Data from Sanche et al. (10)

Notably, while $1/\sigma$, the *mean* latent period, and $1/\gamma$, the *mean* infectious period, are drawn from uniform distributions, under this SE$^m$I$^n$R structure the overall duration of the latent and infectious periods are assumed to be gamma distributed with shape parameters $m$ and $n$ respectively and scale parameters $m\sigma$ and $n\gamma$. The specific values of these distributions were selected based on the literature and recent data.

Sanche et al. drew $M = 10^4$ values for each parameter and accepted those that fell in a serial interval within a range of interest, such that the latent period plus half the infectious period was between 7 and 8 days. The authors then estimate $R_0$ as specified in Equation 8, where $m$ and $n$ are 4.5 and 3. The median $R_0$ as well as a percentile confidence interval, are reported. We reproduced this result, resulting in a median $R_0$ of 5.8 (95% CI: 4.4 - 7.7, R Code 5).

```r
set.seed(8) # set a seed so the result is reproducible
M <- 10^4 # set the number of simulations
lambda <- runif(M, min = 0.21, max = 0.3) # simulate the initial growth rate
gamma <- 1 / runif(M, min = 4, max = 14) # simulate the mean infectious
period
sigma <- 1 / runif(M, min = 2.2, max = 6) # simulate the mean latent period
r0 <- r_seminr(lambda = lambda, gamma = gamma, sigma = sigma,
               m = 4.5, n = 3) # calculate R0

# restrict to serial interval of 7-8 days
serial_interval <- (1 / sigma) + 0.5 * (1 / gamma)
r0 <-   r0[serial_interval >= 7 & serial_interval <= 8]
quantile(r0, c(0.025, 0.5, 0.975)) # calculate the median and 95% CI

##     2.5%      50%    97.5%
##    4.451    5.874    7.747
```

**R Code 5.** Estimating $R_0$ under an SE$^m$I$^n$R model with parameter uncertainty, replicating the Sanche et al. result based on Table 1 inputs.

Even accounting for the uncertainty, the estimate of $R_0$ is highly sensitive to the range of the parameters selected. The Sanche et al. report assumed $\lambda$, the exponential growth, ranged from 0.21 to 0.30 per day (Table 1). Another analysis from Wuhan, China estimated the growth rate, $\lambda$, to be 0.1 (95% CI: 0.05 - 0.16) (14). If instead of using a uniform distribution between 0.21 and 0.30 per day for $\lambda$, we use 0.05 - 0.16, the resulting median $R_0$ is 2.3 (95% CI: 1.6 - 3.3, R Code 6).

```r
set.seed(8) # set a seed so the result is reproducible
M <- 10^4 # set the number of simulations
lambda <- runif(M, min = 0.05, max = 0.16) # simulate initial growth rate
gamma <- 1 / runif(M, min = 4, max = 14) # simulate the mean infectious
period
sigma <- 1 / runif(M, min = 2.2, max = 6) # simulate the mean latent period
r0 <- r_seminr(lambda = lambda, gamma = gamma, sigma = sigma,
               m = 4.5, n = 3) # calculate R0

# restrict to serial interval of 7-8 days
serial_interval <- (1 / sigma) + 0.5 * (1 / gamma)
r0 <-   r0[serial_interval >= 7 & serial_interval <= 8]
quantile(r0, c(0.025, 0.5, 0.975)) # calculate the median and 95% CI

##     2.5%      50%    97.5%
##    1.552    2.294    3.260
```

**R Code 6.** Estimating $R_0$ under an SE$^m$I$^n$R model with parameter uncertainty, using the Sanche et al. inputs for all values except the initial growth rate. Estimating the initial growth rate as ranging between 0.05 and 0.16.

## Refinements for Data Uncertainty

We have shown a simple illustrative example where a known distribution of two parameters can be incorporated into estimates of a third parameter, calculated using a single mathematical relationship between the parameters. Uncertainty in directly specified parameters can also be considered via sensitivity analyses, varying the parameters and examining how the resulting estimates change (6,12). This approach is particularly valuable when more than one seemingly reliable parameter estimate exists and when closed form solutions are not easily derived for the quantities of interest (e.g., when estimating outbreak trajectories).

Often, though, several quantities are to be estimated at once, and from data which do not directly contain information on the parameters of interest. For example, one may wish to estimate the transmission parameter, $\beta$, in multiple cities on the basis of their daily confirmed case rate, or to estimate the age-specific risk of infection from age-stratified prevalence data. In these situations, the model would be fit, or calibrated, to these data (15). There are several ways to evaluate model fit, but all generally involve some iterative process, whereby model results are generated under a given set of parameter values, the model results are compared to data, and the parameter values are updated to achieve a better fit, within some specification for what constitutes good fit (16). There are many statistical methods to quantify model fit, such as weighted least squares (17), approximate Bayesian computation(18), and maximum likelihood-based approaches, including Bayesian MCMC and particle filtering approaches (19–21). In the latter set, the observed data is assumed to follow a distribution with some quantifiable variance, as specified by the likelihood function, in order to calculate uncertainty in the parameters of interest. These methods may also incorporate estimated parameter uncertainty (such as the assumed distribution of the latent period, as discussed above) as informative priors, in Bayesian methods, or as fixed parameters drawn from a known distribution.

Fitting models to data is a valuable tool, particularly for calibrating prediction models with the observed data and for simultaneous inference of multiple quantities. However, not all parameter values are identifiable, particularly when relying on limited or biased data. A key source of data uncertainty is in the observation or reporting process that generates the data available for model fitting. Though it is possible to directly incorporate reporting or observation in model fitting (e.g., as a binomial distribution of the assumed total infections and some reporting or observation probability (22)), this creates an additional parameter that may not be easily separate from, e.g., underlying population susceptibility or trends in transmission (23). Advanced model calibration techniques are helpful but cannot alone resolve inherent uncertainty or bias in data.

# STOCHASTIC UNCERTAINTY

## Stochasticity from Monte Carlo Inference

We consider two primary forms of stochasticity when using transmission models. The first, relevant to our calculation of $R_0$, involves the use of simulations to incorporate uncertainty from known, or estimated, distributions of fixed parameters. In this case, stochasticity in the simulation, or Monte Carlo, process causes variance between simulation runs and is the simplest source of uncertainty to quantify and to reduce (24–26). This *Monte Carlo error* (MCE) can be accounted for by running a simulation process $M$ times and calculating an estimate of the component of model variance due to the Monte Carlo process. A given Monte Carlo simulation consisting of $M$ runs, generates $M$ data samples, $X = \{X_1, X_2, \cdots, X_M\}$. The target quantity of interest, $\phi$, is then estimated as $\hat{\phi}_M(X)$. The MCE is defined as the standard deviation of the Monte Carlo estimator, $\hat{\phi}_M$ (Equation 9).

$$MCE(\hat{\phi}_M) = \sqrt{Var(\hat{\phi}_M)} \qquad (9)$$

This can be estimated using a bootstrap, resampling from $X$ with replacement ($X^*$) and estimating $\hat{\phi}_M(X^*)$. This process can be repeated $B$ times, resulting in $\hat{\phi}_M(X^*_1), \cdots, \hat{\phi}_M(X^*_B)$. The MCE is then estimated as the standard deviation across bootstrapped samples (Equation 10).(26)

$$\widehat{MCE_{boot}}(\widehat{\phi_M}, B) = \sqrt{\frac{1}{B}\sum_{b=1}^{B}\left(\widehat{\phi_M}(X^*_B) - \bar{\widehat{\phi_M}}(X^*)\right)^2} \qquad (10)$$

Since the MCE decreases with increasing sample size(24), in practice, the number of simulations is simply increased until this variance is negligible. The MCE is dependent on both the target quantity of interest, $\phi$, and the underlying data generating mechanism, therefore there is not a rule of thumb for a number of simulation runs, $M$, that is universally appropriate.(26) In our example, Sanche et al. ran their simulation $10^4$ times. This results in a slight variability between runs. We can quantify this variability by bootstrapping this simulation $B$ = 1000 times and examining the distribution of results. Applying this technique to the Sanche et al. estimate of $R_0$, running $10^4$ simulations results in an MCE of 0.035 (Table 2, R Code 7). This can be diminished by increasing the number of simulations, $M$. For example, running this $10^6$ times brings the MCE down to 0.004. Therefore, rather than reporting this type of uncertainty, for this class of mathematical models it is more useful to run a sufficient number of simulations to ensure the estimates are stable and report the number of runs.

**Table 2**. *Stochastic variability in $R_0$ estimation*[a]

| Number of Simulations (M) | MCE of $R_0$ |
|---|---|
| $10^3$ | 0.110 |
| $10^4$ | 0.035 |
| $10^5$ | 0.011 |
| $10^6$ | 0.003 |

M: the number of simulations; MCE: Monte Carlo error
[a]Based on B = 1000 replications

```r
# If you do not have the purrr package, uncomment and run the following:
# install.packages("purrr")
set.seed(8)
M <- 10^4
B <- 1000

r0 <- purrr::map_dbl(1:B, ~ {
  lambda <- runif(M, min = 0.21, max = 0.3)
  gamma <- 1 / runif(M, min = 4, max = 14)
  sigma <- 1 / runif(M, min = 2.2, max = 6)
  r0 <- r_seminr(lambda = lambda, gamma = gamma, sigma = sigma,
                 m = 4.5, n = 3)
  
  serial_interval <- (1 / sigma) + 0.5 * (1 / gamma)
  r0 <- r0[serial_interval >= 7 & serial_interval <= 8]
  median(r0)
})

(mce <- sd(r0))

## [1] 0.0349
```

**R Code 7**. Estimating the Monte Carlo error (MCE) for the Sanche et al. $R_0$ estimate. The second general form of stochasticity in transmission models

## Process Stochasticity in Disease Transmission

The form of stochasticity described above considers observation stochasticity, or the idea that available parameter values come from an observation of the true parameter. These incomplete observations of the truth produce uncertainty in the parameters used to estimate the quantity of interest but, through simulation, we can propagate this parameter uncertainty to estimate a distribution of the quantity of choice.

The second form of stochasticity, by contrast, considers the inherent process stochasticity of disease transmission. At any moment, whether a given transition event (e.g. infection) occurs is a factor both of the defined model quantities (e.g., the transmission coefficient, $\beta$, and the current prevalence of infectious individuals, $I$) and a series of complex, random processes (e.g., person-person or person-vector interactions; biological infection processes). Note that we consider these as random processes within the context of our model; while such processes may theoretically be quantifiable, the complexity required to model their occurrence far surpasses the scope of disease transmission models and most data.

These stochastic effects limit the potential accuracy of model forecasts, even when much is known about underlying dynamics and drivers of disease transmission, particularly at the beginning or end of an outbreak when the occurrence (or non-occurrence) of any single transition can have an outsized effect on outbreak trajectories (15). Thus, models seeking to predict outcomes, such as the final outbreak size, or the number of hospitalizations expected in a given time, must account for uncertainty generated by this stochasticity. In practice, this often takes the form of random, binomial draws at each time step, of size $\Delta t$, to determine the number of individuals undergoing each transition. In an SEIR model, this includes the number newly infected ($X_t$), newly infectious ($Y_t$), and newly recovered ($Z_t$) at each time step. The probability of each transition is derived from the system of ODEs, where the probability of a transition occurring is equal to $1 - exp(-\phi \cdot \Delta t)$, if $\phi$ is the rate of leaving the originating compartment (Equation 11). R Code 8 demonstrates how to perform this simulation; there are several R packages that can also be used for this purpose, including **EpiModel**, **SimInf**, and **RLadyBug** (27-29).

$$X_t \sim Binom(S_t, 1 - exp(-\beta \, I_t \, \Delta t)) \tag{11}$$

$$Y_t \sim Binom(E_t, 1 - exp(-\sigma \Delta t))$$

$$Z_t \sim Binom(I_t, 1 - exp(-\gamma \Delta t))$$

$$S_{t+1} = S_t - X_t$$

$$E_{t+1} = E_t + X_t - Y_t$$

$$I_{t+1} = I_t + Y_t - Z_t$$

$$R_{t+1} = R_t + Z_t$$

```r
set.seed(8)
simulate_seir <- function(beta, sigma, gamma, initial_cond, delta_t = 0.25) {
  t <- 0 # set the first time point (in days)
  # set output matrix
  out <- data.frame(t = t,
                    S = initial_cond[["S"]],
                    E = initial_cond[["E"]],
                    I = initial_cond[["I"]],
                    R = initial_cond[["R"]])
  y  <- initial_cond # set y, temporary variable
  while (y["I"] > 0) {
    t <- t + delta_t
    # define the probability of infection
    p_infect <- 1 - exp(-delta_t * beta * y["I"])
    
    # define the probability of leaving the latent period
    p_latent <- 1 - exp(-delta_t * sigma)
    
    # define the probability of recovering (from infectious)
    p_recover <- 1 - exp(-delta_t * gamma)
    
    # randomly draw the number newly infected
    incident_cases <- rbinom(1, y["S"], p_infect)
    
    # randomly draw the number newly infectious
    infectious_cases <- rbinom(1, y["E"], p_latent)
    
    # randomly draw the number newly recovered
    recovered_cases <- rbinom(1, y["I"], p_recover)
    y <- c(S = y[["S"]] - incident_cases,
           E = y[["E"]] + incident_cases - infectious_cases,
           I = y[["I"]] + infectious_cases - recovered_cases,
           R = y[["R"]] + recovered_cases)
    out <- rbind(out, c(t, y))
  }
  return(out)
}

d <- simulate_seir(0.001, 1/5, 1/5,
                   initial_cond = c(S = 999, E = 0, I = 1, R = 0))
head(d)

##      t   S E I R
## 1 0.00 999 0 1 0
## 2 0.25 999 0 1 0
## 3 0.50 998 1 1 0
## 4 0.75 998 1 1 0
## 5 1.00 997 2 1 0
## 6 1.25 997 2 1 0
```

**R Code 8**. Simulating a disease outbreak using an SEIR model. This code includes a function to perform the simulation as well as an example where $\beta = 0.001$, $\sigma = 1/5$, $\gamma = 1/5$, and the initial values for the number of susceptible, exposed, infected, and recovered individuals is 999, 0, 1, and 0, respectively. The function outputs a data frame with the estimated number of individuals in each compartment at each time. We show the first 6 rows of output, corresponding to times $t = 0$ to $t = 1.25$.

Unlike the simulation stochasticity described above, in which a distribution of estimates is derived from a distribution of parameters, process stochasticity produces a distribution of estimates for a single parameter set. The goal here is not to reduce this uncertainty, but instead to consider a sufficient number of simulation results to properly characterize the range of possible model outcomes given randomness of disease transmission.

# STRUCTURAL UNCERTAINTY

The final type of uncertainty is *structural uncertainty*. Structural uncertainty refers to uncertainty in the model choice (30,31). In Sanche et al, the model structure is assumed to be an $SE^mI^nR$ model where latent and infectious time is gamma distributed. There are other potential structures that could have been used to explain this relationship (though more complex models will generally introduce greater data uncertainty, as more parameters must be input or inferred (31)). Choice of structure is an important assumption in all modeling analyses, and these assumptions should be clearly grounded in the available evidence. Early in an outbreak of an emerging disease, there may be greater uncertainty in what constitutes an appropriate model; for example, whether there is a latent period or certain high-risk groups that should be modeled separately. As an outbreak progresses, though, it becomes more clear which features are most important for disease transmission and therefore which structures are appropriate. For example, for most applications for COVID-19, the SEIR model structure is preferable to SIR, as the SIR simplification underestimates the impact of this particular epidemic, specifically due to the lack of an "Exposed" compartment.

Uncertainty in this model choice is difficult to characterize, however some sensitivity analyses could be implemented. For example, we can estimate what $R_0$ would be under SIR or SEIR models, rather than the more flexible SE$^m$I$^n$R. Figure 4 demonstrates the different distribution of estimated $R_0$ values depending on the chosen structure SIR, SEIR, or SE$^m$I$^n$R based on the parameters defined in Equation 1, Equation 4, and Equation 7 and a serial interval of 7-8 days (R Code 9).

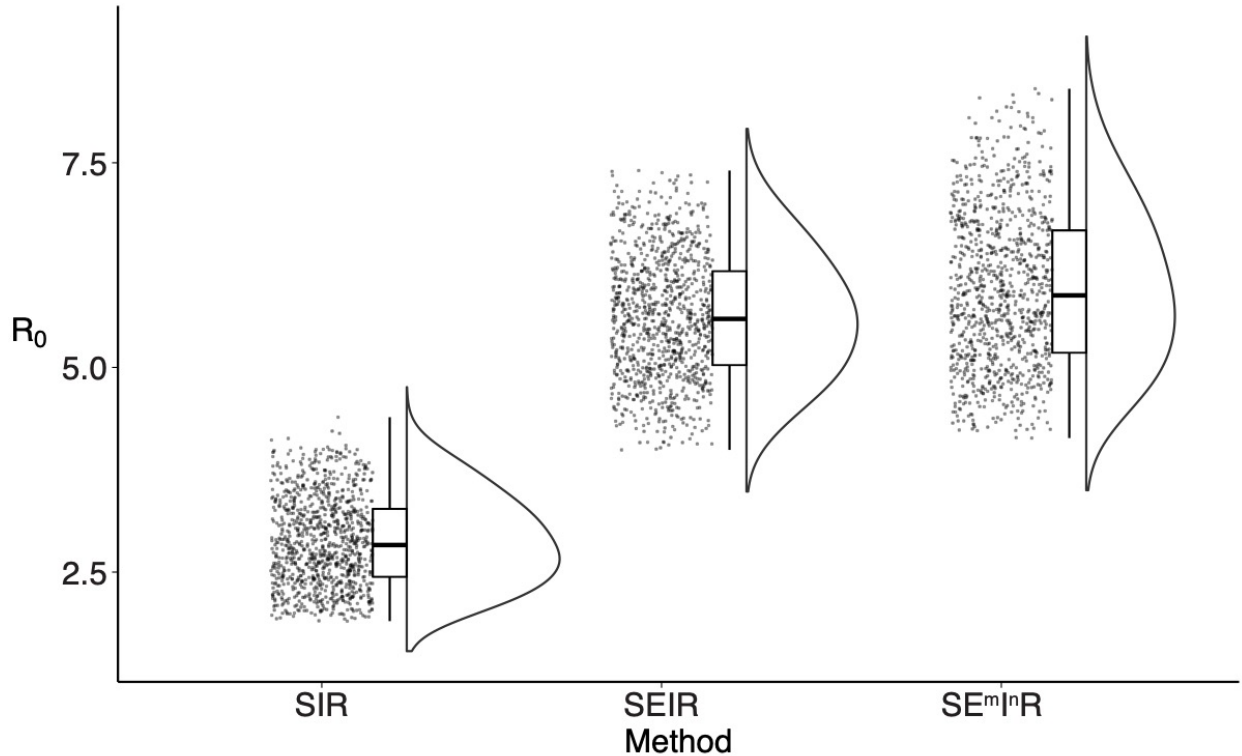

**Figure** 4. Estimated $R_0$ depending on the specified model structure.

```r
set.seed(8)
M <- 10^6
lambda <- runif(M, min = 0.21, max = 0.3)
gamma <- 1 / runif(M, min = 4, max = 14)
sigma <- 1 / runif(M, min = 2.2, max = 6)
serial_interval <- (1 / sigma) + 0.5 * (1 / gamma)

r0_sir <- r_sir(lambda = lambda, gamma = gamma)
r0_seir <- r_seir(lambda = lambda, gamma = gamma, sigma = sigma)
r0_seminr <- r_seminr(lambda = lambda, gamma = gamma, sigma = sigma,
                      m = 4.5, n = 3)
r0_sir <- r0_sir[serial_interval >= 7 & serial_interval <= 8]
r0_seir <- r0_seir[serial_interval >= 7 & serial_interval <= 8]
r0_seminr <- r0_seminr[serial_interval >= 7 & serial_interval <= 8]

quantile(r0_sir, c(0.025, 0.5, 0.975))

##     2.5%     50%    97.5%
##    2.002   2.840   3.925

quantile(r0_seir, c(0.025, 0.5, 0.975))

##     2.5%     50%    97.5%
##    4.305   5.552   7.052

quantile(r0_seminr, c(0.025, 0.5, 0.975))

##     2.5%     50%    97.5%
##    4.407   5.826   7.846
```

**R Code 9.** Calculate an estimated $R_0$ with 95% CIs under SIR, SEIR, and $SE^mI^nR$ model, assuming the Sanche et al. parameter values specified in Table 1.

The SIR model results in a median $R_0$ of 2.8 (95% CI: 2.0 - 3.9), the SEIR model results in a median $R_0$ of 5.5 (95% CI: 4.3 - 7.1), and the $SE^mI^nR$ model assuming gamma distributions results in a median $R_0$ of 5.8 (95% CI: 4.4 - 7.8).

Another approach to characterizing structural uncertainty is to consider the underlying causal model represented by our mechanistic model. Ackley et al. provides a mathematical framework for mapping between these two representations.(32) By describing the mechanistic model in this way, we can interrogate assumptions about sources of parameter values and relationships between modeled compartments.

In the directed acyclic graph, common causes of infectious contacts and disease progression provides a potential source of bias in the estimation of the relationship between infectious contact, infection, and recovery (Figure 5). If these potential confounders differ between the population from which parameter data were obtained (here, Wuhan, China) and the population for which inference is desired (here, USA) then the model inference can be biased.(33) A full assessment of structural uncertainty thus requires consideration of the causal relationships between modeled variables within the model and in the populations of interest.

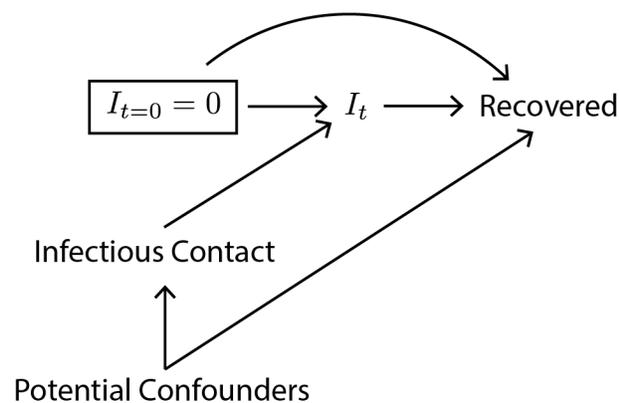

**Figure** 5. One possible directed acyclic graph representation of the SEIR compartment model. $I_{t=0}$ represents infection status at baseline and is restricted to uninfected (0) individuals. Infectious contacts are required for infection status ($I_t$) to change over time, and infection status and time both determine recovery status.

A final source of structural uncertainty arises from scenarios under which model results are obtained, when the mathematical model is being used to predict future disease trajectories. All outbreaks, but particularly the COVID-19 pandemic, are marked by frequent changes to available and implemented interventions. Assumptions about which interventions are in place at future timepoints in a transmission model impacts predictions tremendously. Moreover, assumptions about relationships between intervention variables and model components can lead to bias when attempting to infer intervention effectiveness. When assessment of an intervention is desired, use of directed acyclic graphs to assess the model structure is highly recommended.

## DISCUSSION

Mechanistic models are an important tool for understanding complex systems, but use of mechanistic models for infectious disease transmission does not exempt us from the need to address statistical uncertainty. Any mechanistic model which reports only a single result should be viewed with the same skepticism as a statistical analysis which reports only the point estimate without information on the variance. At minimum, mechanistic models should explicitly state potential sources and magnitudes of uncertainty, as described here. We have outlined both simple and more advanced techniques to do so.

It is worth noting that data used to inform transmission models, especially early in an epidemic, are likely to be biased. Among others, time-to-event distributions will likely be incompletely observed, and changes in reporting completeness may be indistinguishable from changes in transmission; these biases will propagate to other model estimates. Advanced techniques alone cannot resolve this issue, but during epidemics the direction of the bias may be predictable allowing for thoughtful implementation of sensitivity analyses (34). As an epidemic continues, the infectious disease modeling community often gains knowledge about the appropriate structure for the underlying model, and a better understanding of the quality of available data. The use of ensemble methods and standardized reporting guidelines can facilitate exploration of structural uncertainty and assumptions across multiple models (35,36).

Mechanistic models rely on our best current understanding of the system and results of models should be presented in the context of what is known and unknown at the time of modeling. It should be expected and desired that models will change in structure and parameterization over time, becoming better fit to the data as more data are available. As George Box famously noted, all models are wrong but some are useful.(1,37) Importantly, the usefulness of a model is strongly tied to how well the assumptions on which it rests have been explained and how clearly the uncertainty has been specified.